\documentclass[letterpaper]{article} 
\usepackage{aaai2026}  
\usepackage{times}  
\usepackage{helvet}  
\usepackage{courier}  
\usepackage[hyphens]{url}  
\usepackage{graphicx} 
\usepackage{natbib}  
\usepackage{caption} 
\usepackage{algorithm}
\usepackage{algorithmic}
\usepackage{amsmath}
\usepackage{amsfonts}
\usepackage{amssymb}
\usepackage{xcolor}
\usepackage{newfloat}
\usepackage{listings}
\usepackage{booktabs}
\usepackage{siunitx}
\usepackage{makecell} %
\usepackage{subfigure}

\DeclareCaptionStyle{ruled}{labelfont=normalfont,labelsep=colon,strut=off} 
\floatstyle{ruled}
\newfloat{listing}{tb}{lst}{}
\floatname{listing}{Listing}
\title{Federated Inference for Heterogeneous LLM Communication and Collaboration}
\author {
    Zihan Chen\textsuperscript{\rm 1}\thanks{Equal Contributions. This work was supported in part by SUTD Kickstarter Initiative (SKI 2021\_06\_08), and in part by the National Research Foundation, Singapore and Infocomm Media Development Authority under its Communications and Connectivity Bridging Funding Initiative. (\textit{Corresponding authors are J. Park and H. H. Yang})},
    Zeshen Li\textsuperscript{\rm 2}$^*$,
    Howard H. Yang\textsuperscript{\rm 2},
    Tony Q.S. Quek\textsuperscript{\rm 1},
    Jihong Park\textsuperscript{\rm 1}
}
\affiliations {
    \textsuperscript{\rm 1}Singapore University of Technology and Design, Singapore\\
    \textsuperscript{\rm 2}ZJU-UIUC Institute, Zhejiang University, Haining, China\\
    zihan\_chen@sutd.edu.sg, zeshen.22@intl.zju.edu.cn, haoyang@intl.zju.edu.cn, \\ tonyquek@sutd.edu.sg, jihong\_park@sutd.edu.sg
}

\usepackage{bibentry}

\begin{document}

\maketitle

\begin{abstract}
Given the limited performance and efficiency of on-device Large Language Models (LLMs), the collaborations between multiple LLMs enable desirable performance enhancements, in which data, tokens, and model weights could be shared across LLMs. This process is constrained by task-oriented QoS demands, privacy requirements, and inherent system heterogeneity. In view of the above challenge and to fully exploit the on-device inference capabilities, we present a novel federated inference framework in this position paper, termed federated refinement \texttt{FedRefine}. This framework presents a new paradigm for heterogeneous LLMs collaboratively performing inference with communicating KV caches in a privacy-preserving manner. Some numerical results are provided to highlight the superiority of \texttt{FedRefine}. Several interesting topics are also highlighted for future research. By exploring the LLM-native communications, we wish to provide a new paradigm for this broad area. 
\end{abstract}

\section{Motivation}
Recent advances have enabled edge devices to host small and large language models (LLMs) locally~\cite{yang2025qwen3,friha2024llm}. However, these on-device models still suffer from compromised inference accuracy and speed compared to full-scale cloud LLMs. Fully offloading inference to the cloud by transmitting all input and output tokens is certainly not scalable, as it entirely neglects the potential of on-device inference capabilities. Alternatively, motivated by federated learning~\cite{MaMMooRam:17AISTATS}, which utilizes local computation and data for collaborative training, this work aims to put forward a new direction of federated inference that collaboratively exploits on-device inference capabilities to achieve fast and accurate results collectively. 

Achieving this goal is non-trivial due to challenges in inference latency, privacy, and heterogeneity as follows. Specifically, in modern autoregressive LLMs, each output token is generated not immediately after decoding the last token but after decoding all its previous tokens to ensure contextual consistency. Therefore, inter-device token communication such as text-to-text communication (T2T), induces significant computation latency, equivalent to the pre-fill delay required to rebuild the key-value (KV) cache. Furthermore, LLM input and output tokens are often human-interpretable and may reveal private user content. Also, the heterogeneous nature of model architectures restricts the exchange of architecture-dependent semantic information and knowledge required for LLM communication and collaboration.
To address these challenges, we propose a novel federated inference framework, termed federated refinement, \texttt{FedRefine}, where devices communicate KV caches instead of tokens, thereby skipping pre-fill delays and generating a large number of new tokens while keeping private tokens locally. \texttt{FedRefine} is built upon two key ideas, LLM self-refinement (SelfRefine) and cache-to-cache communication (C2C), as elaborated in the following section.

\section{Rethinking Heterogeneous LLM Communications and Collaborations}\label{sec:main}

Building upon the principles of SelfRefine and C2C, this section presents \texttt{FedRefine} to enable efficient collaborative inference across heterogeneous LLMs.

\subsection{From Self-Refine to Cache-to-Cache (C2C)}
\begin{figure}[t!]
\vspace{1cm}
  \centering
  \includegraphics[width=0.85\columnwidth]{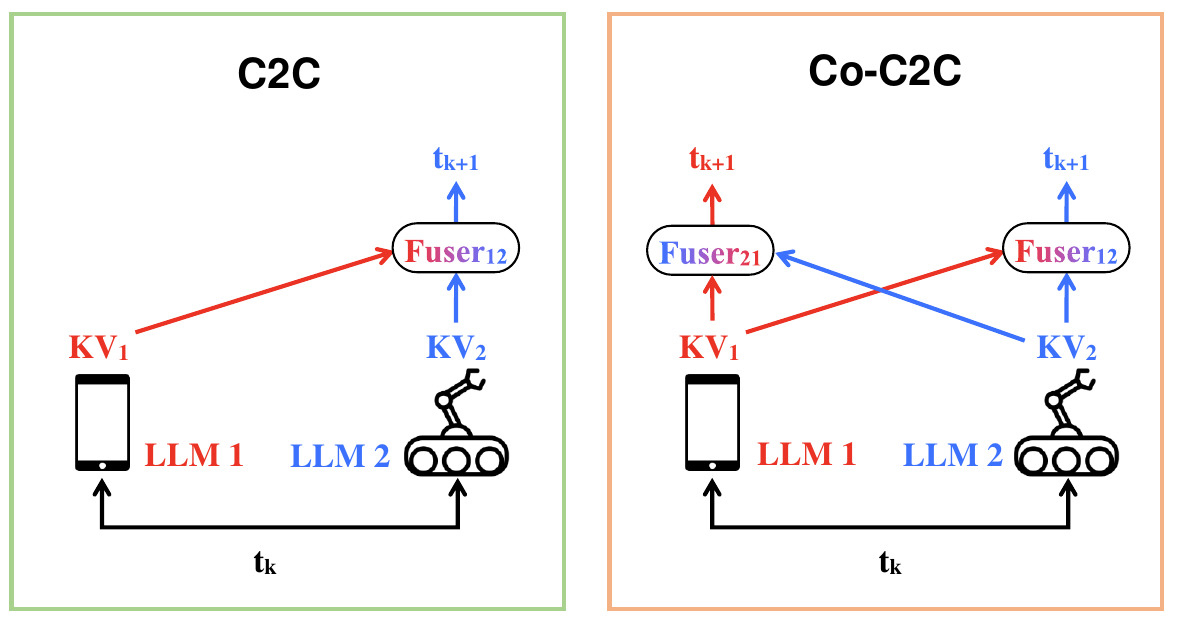}
  \caption{Illustration of unidirectional and bidirectional cache communication.}
  \label{fig:sys1}
\end{figure}

To exploit the inference capabilities of on-device models, self-refinement allows LLM to iteratively improve its own output~\cite{madaan2023self}.
However, this on-device refinement process is fundamentally limited by the model's internal knowledge. To overcome this limitation and enable collaborative refinement for improved inference, models must exchange information. In such context, the device performing the primary inference is termed the receiver, while the device sharing data or knowledge to assist is termed the transmitter. 
Intuitively, such collaboration occurs at the output level via T2T communication, where LLMs communicate with natural language tokens. This approach improves inference performance but incurs significant latency overhead. 
Subsequently, a more efficient approach via C2C communication was proposed~\cite{fu2025c2c}, mitigating the high latency issues associated with T2T, in which the LLM at the receiver (i.e., LLM 2) performs refined inference by leveraging the KV Cache of the transmitter's model (i.e., LLM 1)  as shown in Fig.~\ref{fig:sys1}(left), in which both LLMs share the same input $\mathbf{t}_k$. 
In particular, the bridge across the KV cache from LLM 1 to LLM 2 could be implemented via a pre-trained C2C fuser (i.e., $\text{Fuser}_{12}$ in Fig.~\ref{fig:sys1}(left)). 
Denote the LLM 1 at the transmitter, KV Cache, and the corresponding $\text{Fuser}_{12}$ as $\mathcal{M}_1$, $\mathcal{C}$, and $\mathcal{F}_{12}$, respectively. With the model $\mathcal{M}_2$ (i.e., LLM 2) and the current token $t_k$, the next token predicted in the decoding process at LLM 2 is given by 
\begin{equation}\label{eq:uni_c2c}
    t_{k+1}=\mathcal{P}(t_k|\mathcal{C}(\mathcal{F}_{12},\mathcal{M}_1)\circ \mathcal{C}(\mathcal{M}_2)),
\end{equation}
in which $\circ$ is the sequence-wise concatenation operation. 
Such a framework allows for the unidirectional communication of internal KV cache states with richer semantic knowledge compared to text, enabling far more efficient collaboration and inference refinement while avoiding the high latency and information loss of text-based interactions.

\subsection{From Unidirectional C2C to Bidirectional Co-C2C}
The C2C framework defines a unidirectional collaborative inference scenario. 
Analogous to the co-distillation approach, we introduce a bidirectional variant of the C2C scheme. As shown in  Fig.~\ref{fig:sys1}(right), the core of this approach involves training two fusers (i.e., $\text{Fuser}_{21}$ and $\text{Fuser}_{12}$ for LLM 1 and LLM 2, respectively) to bidirectionally bridge any two LLMs, thereby facilitating their communication and collaboration for refined inference performance. In addition to the decoding process given in Eq.~\ref{eq:uni_c2c}, a reverse refined decoding process could be obtained via
\begin{equation}\label{eq:uni_c2c}
    t_{k+1}=\mathcal{P}(t_k|\mathcal{C}(\mathcal{F}_{21},\mathcal{M}_2)\circ \mathcal{C}(\mathcal{M}_1)),
\end{equation}
in which $\mathcal{F}_{21}$ represents the fuser for projecting KV cache from LLM 2 to LLM 1. 
Overall, bidirectional cache communication surpasses one-way updates by enabling mutual refinement, fostering a fairer and incentive-compatible collaboration paradigm, by allowing devices to assume dual roles, acting simultaneously as receiver and transmitter.

\subsection{Fed-Refine: Federated Inference with Refinement}

\begin{figure}
  \centering
  \includegraphics[width=0.78\columnwidth]{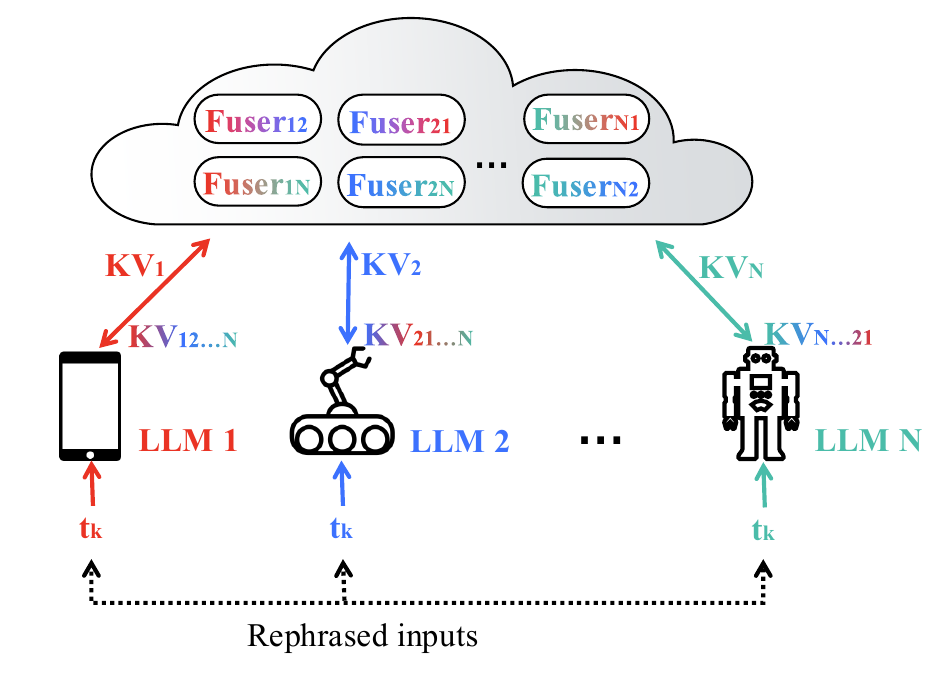}
  \caption{A depiction of the federated refinement framework in a heterogeneous multi-LLM system.}
  \label{fig:sys2}
\end{figure}

Built upon the bidirectional cache communication-based collaboration, we now formally present \textit{Federated Refinement} framework, \texttt{FedRefine}, as depicted in Fig.~\ref{fig:sys2}.
In particular, our proposed \texttt{FedRefine} integrates Cache communication as the medium of collaboration over the heterogeneous multi-LLM systems, without requiring identical model architectures.  
By maintaining all KV cache fusers for all possible bidirectional collaborations, \texttt{FedRefine} enables a model-agnostic and bidirectional KV Cache sharing-based collaboration paradigm across heterogeneous LLMs, where LLMs will perform inference with rephrased input tokens to ensure privacy protection without any intent leakage. 

Take Fig.~\ref{fig:sys2} as an example. Consider a system with $N$ LLMs, in which all the inference computation will be collaboratively accomplished via sharing corresponding KV Cache for any certain task $\mathcal{T}$.
The server maintains all pre-trained fusers, $\{\mathcal{F}_{12},\mathcal{F}_{21}, \mathcal{F}_{ij}, \mathcal{F}_{ji}, \dots, \mathcal{F}_{1N}, \mathcal{F}_{N1}\}$, in which $\mathcal{F}_{ij}$ and $\mathcal{F}_{ji}$ represent the fuser pair for any bidirectional cache communication link, $i\leftrightarrow j$, in which $i, j\in\mathcal{N}$. The pre-training of each fuser are conducted separately for each pair of LLM collaboration and the training process could refer~\cite{fu2025c2c}.
In addition, a gating network is required for each LLM to select the data from its own model or other fusers.
We now introduce how \texttt{FedRefine} works from two LLM collaborations and multiple LLM collaborations.  At the beginning of inference, the LLMs at both the transmitter and receiver receive distinct rephrased input tokens to preserve privacy. For simplicity, we let $t_k$ denote the rephrased tokens for all LLMs when performing the $k$-th inference task.
In \texttt{FedRefine}, for any selected two LLMs $i$ and $j$ with input tokens $t_k$, the bidirectional cache communication-based refined inference is given by
\begin{align}
   \small t_{k+1}=\mathcal{P}_j(t_k|\mathcal{C}(\mathcal{F}_{ij},\mathcal{M}_i)\circ \mathcal{C}(\mathcal{M}_j)), \text{for LLM } i \rightarrow j; \nonumber\\
   \small t_{k+1}=\mathcal{P}_i(t_k|\mathcal{C}(\mathcal{F}_{ji},\mathcal{M}_j)\circ \mathcal{C}(\mathcal{M}_i)), \text{for LLM } j \rightarrow i. 
\end{align}\label{eq:uni_c2c}
    
\begin{figure*}[t]
    \centering
    \subfigure[Federated Inference Accuracy]{
        \includegraphics[width=0.3\textwidth]{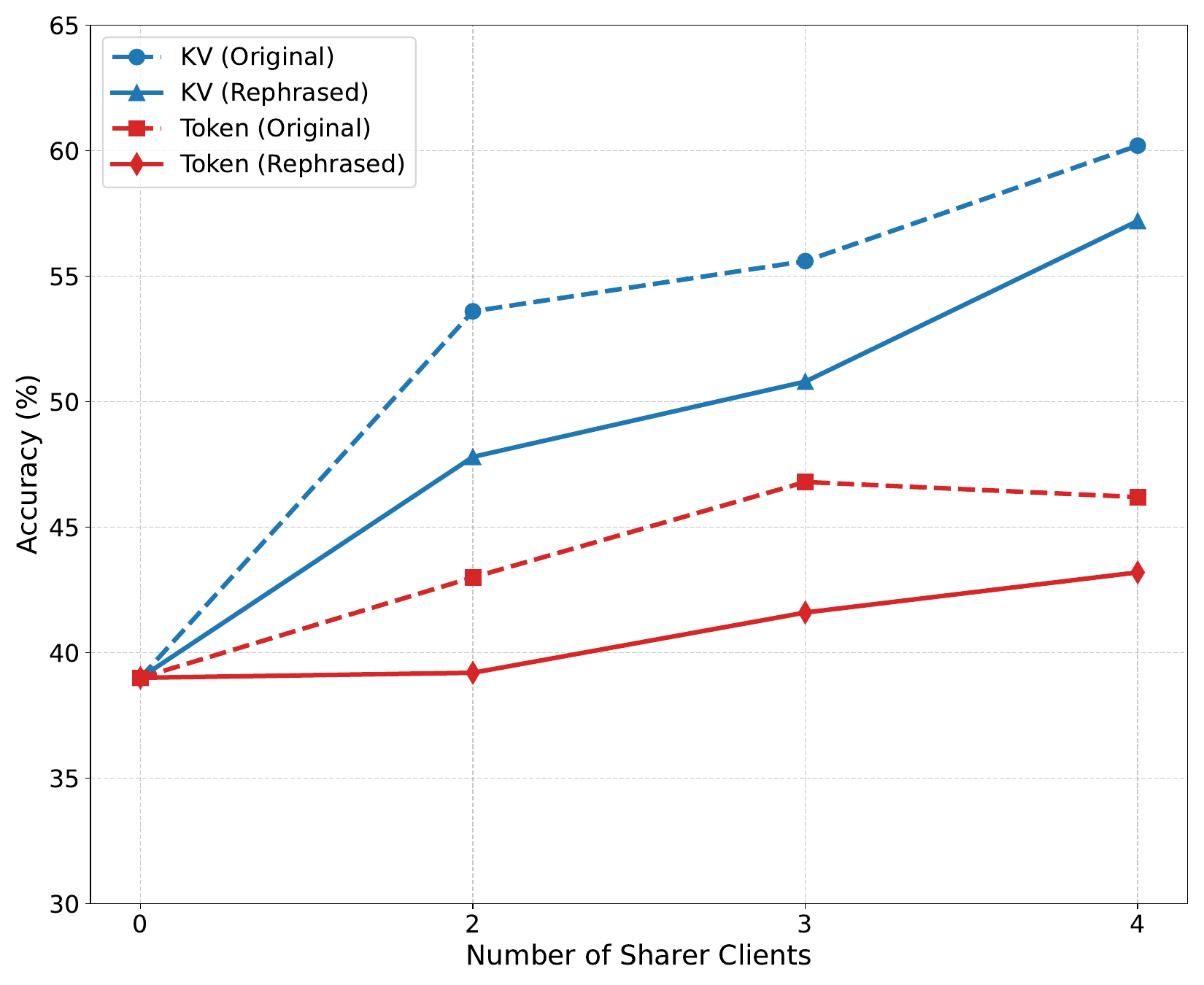}
        \label{fig:persubfig1}
    }
    \hspace{1em} 
    \subfigure[Point-to-Point C2C/T2T]{
        \includegraphics[width=0.3\textwidth]{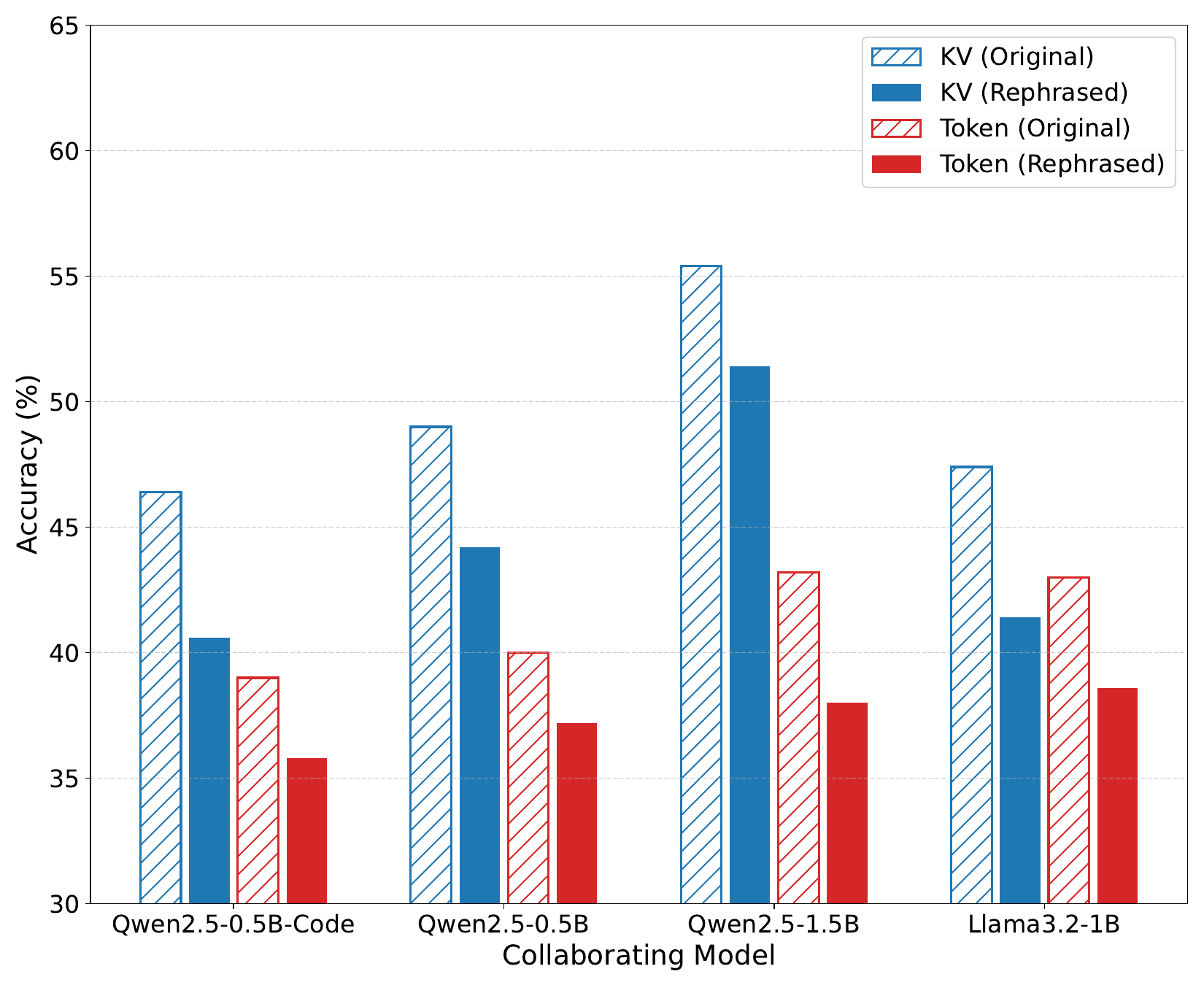}
        \label{fig:persubfig2}
    }
    \hspace{1em}
    \subfigure[Federated Inference Latency]{
        \includegraphics[width=0.3\textwidth]{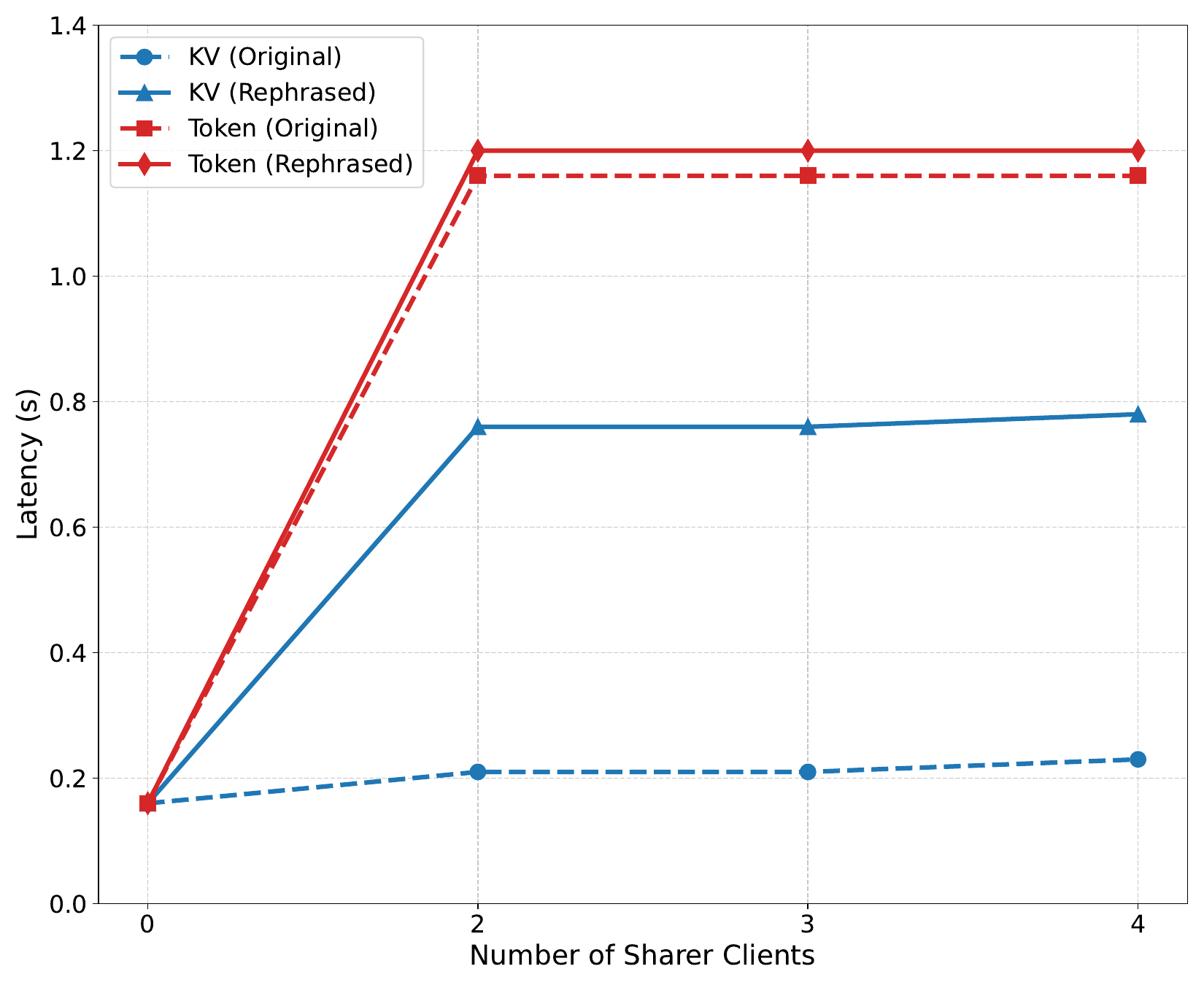}
        \label{fig:persubfig3}
    }
    \caption{Performance evaluation of the proposed collaborative inference framework. ``KV'' and ``Token'' denote collaborative protocols via C2C and T2T transmission, respectively. ``Original'' refers to transmitting raw queries without privacy protection, while ``Rephrased'' indicates the use of privacy-preserving semantically rewritten queries.}
    \label{fig:per}
\end{figure*}

When extended to multiple LLM cache communication scenarios, the refined  inference process could be given by 
\begin{align}\label{eq:uni_c2c}
    t_{k+1}&=\mathcal{P}_i(t_k|\mathcal{C}(\mathcal{F}_{j_1i}, \mathcal{M}_{j_1}) \circ \nonumber \\
   &\mathcal{C}(\mathcal{F}_{j_2i},\mathcal{M}_{j_2})\circ \dots\circ \mathcal{C}(\mathcal{F}_{j_si},\mathcal{M}_{j_s})\circ\mathcal{C}(\mathcal{M}_i))
\end{align}
in which LLM $i$ leverage the KV cache from multiple heterogeneous LLM $j_1, j_2, \dots, j_s$. 
In summary, via implementing the multi-bidirectional KV cache fuser, \texttt{FedRefine} enables flexible cache communication and scalable collaborative LLM inference over heterogeneous networks.

\section{Case Study}
To further facilitate the above analysis of our proposed \textit{FedRefine}, we provide a case study on evaluating the collaboration performance over a heterogeneous multi-LLM system, which comprises a receiver model, Qwen3-0.6B~\cite{yang2025qwen3}, and four transmitter models: Qwen2.5-0.5B, Qwen2.5-0.5B-code, Qwen2.5-1.5B~\cite{hui2024qwen2}, and Llama-3.2-1B~\cite{dubey2024llama}.
The system schedules different pairs of LLMs to perform collaborative inference through cache communications, with the server hosting a total of four distinct fusers for KV Cache projection across these pairs. Motivated by~\cite{fu2025c2c}, we align the receiver and sender models layer-by-layer from the bottom up. For each layer of the receiver, a three-layer MLP network projects the corresponding layer’s KV Cache from the sender model into the receiver model. The receiver then mixes the projected KV Cache with its own as the updated KV Cache. Each C2C fuser is trained on the first 50,000 samples of the general fine-tuning dataset OpenHermes2.5~\cite{teknium2023openhermes}, and evaluation of detection performance is conducted using the OpenBookQA dataset~\cite{mihaylov2018can}. In this paper, we employ the receiver model, Qwen3-0.6B, to rephrase the original questions. We use the average inference time as the latency evaluation metric (in seconds), and measure the communication load by the amount of data transmitted per token in both text transmission and cache communications scenarios.

Based on the aforementioned fuser network architecture, we conducted performance evaluations for standalone model inference and heterogeneous LLM federated inference with the results shown in Fig.~\ref{fig:per}. Specifically, the heterogeneous LLM federated inference experiments were designed based on prior knowledge that different models exhibit varying performance across different tasks. Based on this prior knowledge, the receiver model selects different model combinations according to the different tasks.
First, as depicted in Fig.~\ref{fig:persubfig1}, all federated inference models surpass the receiving model's standalone baseline, with accuracy showing a clear upward trend as more sharing models participate. This confirms the effectiveness of our LLM collaborative inference framework.
For instance, with all four sharing models participating, the non-private KV collaborative model yields a 21.2\% accuracy improvement over the independent inference baseline. Furthermore, the privacy-preserving KV model exhibits only a 3\% decrease in accuracy under the same setting, demonstrating that our privacy protection strategy does not significantly compromise collaborative performance.
Finally, the accuracy of federated inference via C2C significantly outperforms the T2T approach. Specifically, with the full participation of four sharer models, C2C surpasses T2T by approximately 15\%. However, in this full-participation setting, transmitting the KV cache for a single token requires 88 KB, whereas T2T requires only 16 bytes, indicating that the C2C approach imposes significantly higher demands on communication resources.
Fig.~\ref{fig:persubfig2} indicates that the intrinsic capabilities of the sharer model directly impact the performance of the collaborative model. Meanwhile, Fig.~\ref{fig:persubfig3} reveals that although the privacy-preserving C2C method incurs additional latency due to query rewriting, its total latency remains significantly lower than that of the T2T approach.
Overall, results show the proposed heterogeneous LLM federated inference paradigm achieves efficient knowledge transfer and significant performance gains with low costs.

\section{Possible Variants and Future Trends}
\label{sec:future}
Considering the challenges brought by the overhead of the exchanged cache data, the decision to use cache or token communication could be dynamically determined based on both the current network status and the specific QoS requirements. 
That said, cache communication and token communication could be adopted in an opportunistic manner.
The shared tokens and KV Cache data undergo processing on the edge server (e.g., via a prompt summarizer/optimizer~\cite{yuksekgonul2024textgrad} or KV Cache fuser), enabling target clients to leverage the processed information for further collaborative task execution, thereby enhancing overall inference performance.

Building upon the \texttt{FedRefine} paradigm, we now outline key challenges and future research directions for achieving sustainable foundation model collaborations for next-generation intelligent networks. 

\begin{itemize}
    \item \textbf{Iterative local refinement.} Existing self-refinement methods typically involve a single LLM that iteratively generates and revises its own output to enhance response quality without external knowledge sharing or supervision~\cite{madaan2023self}. How to effectively design iterative inference and refinement with cache or token communications could be further explored. 
    \item \textbf{Continuous global federation iterations.} The potential for a single model's refinement to enhance its collaborators motivates a new research direction. We can mirror the concept of local iterative refinement in a collaborative setting, exploring multi-iteration cache communication as a mechanism to achieve continuous, system-wide LLM refinement.
    \item \textbf{Cache communication for multi-modal LLMs.} The evolution from text-only LLMs to multi-modal models unlocks a much wider range of applications but also introduces more diverse challenges~\cite{hu2024bliva}, making it essential to design effective cache communication strategies tailored for these multi-modal scenarios.
    \item \textbf{Prompt engineering for federated inference.} In LLM collaborations, prompt engineering is crucial for defining each model's specific role and orchestrating their interaction to achieve a unified goal~\cite{white2023prompt}. In the context of federated inference, it would be necessary to develop prompt engineering approaches for cache communication-based model refinement in a privacy-preserving manner.   
\end{itemize}

\section{Conclusion}

This work offers diverse perspectives on LLM communication collaborations. Inspired by the C2C approach, we proposed \textit{Federated Refinement}, a novel framework that leverages bidirectional cache communication to achieve scalable collaborative inference between heterogeneous LLMs. By exploring the LLM-native communications with validated efficacy, we wish to provide a new paradigm for this broad area. 

\bibliography{aaai2026}
\newpage

\appendix
\section{Appendix-A: Extended version of \texttt{FedRefine}}
With different network connections and heterogeneity status, our proposed framework evolves into following major types.
\begin{enumerate}
    \item \textit{Homogeneous} federated inference and refinement via  \textit{token communication}; 
    \item  \textit{Homogeneous} federated inference and refinement via  \textit{cache communication};
    \item  \textit{Heterogeneous} federated inference and refinement via  \textit{token communication};
    \item  \textit{Heterogeneous} federated inference and refinement via  \textit{cache communication}.
\end{enumerate}

\section{Appendix-B: Literature Review}

To address the limitation of single models, LLM collaboration has emerged as a promising alternative that distributes the computational and cognitive burden across multiple models rather than relying on a single architecture~\cite{li2023camel}. In collaborative frameworks, heterogeneous LLMs, which could be deployed across multiple devices,  edge or cloud server, to handle complex queries by leveraging their computational capacity-dependent complementary strengths. Concretely, consider a multi-LLM system deployed over a heterogeneous and resource-constrained edge network, in which on-device LLMs are maintained according to each device's maximum computational capacity~\cite{friha2024llm}. Multiple LLMs with different architectures and scales would be selected by the edge server to communicate for collaborative tasks, such as hybrid inference, knowledge transfer, model refinement, or joint model optimization and fine-tuning~
\cite{zhang2024edgeshard}. 
Such collaborative schemes exhibit varying communication and computational overhead profiles, distinguished by the nature of transmitted information. Communication payloads may consist of discrete tokens for prompt-based interactions, gradient tensors for backpropagation-based learning, complete or partial model weight matrices for direct parameter sharing, or intermediate embeddings for feature-level knowledge exchange.

Therefore, the above constraints motivate our exploration of an adaptive, task-oriented, and model-agnostic framework for LLM-native communication and collaboration. More specifically, while tokens, gradients, and model weights are common mediums for designing LLM communication and collaboration frameworks, the KV Cache, the core and LLM-native representation of knowledge, has been largely overlooked~\cite{fu2025c2c}. 
On the other hand, either in hybrid inference or knowledge distillation, a larger model is always assumed to be capable of rectifying or improving less powerful model's output~\cite{naveed2025comprehensive}. It would be also interesting to explore the reverse way of communication flow.

The existing LLM collaboration solutions are mainly built upon the adaptive task scheduling and routing strategies, whereby the system can adjust its collaboration policy based on real-time and task-oriented factors such as query complexity, available resources, network conditions, and latency requirements~\cite{li2024eagle}. By scheduling multiple LLMs through intelligent task decomposition, selective offloading, and result synthesis, collaborative frameworks can achieve performance levels approaching or exceeding those of larger monolithic models while maintaining the efficiency and privacy benefits critical for on-device deployment.
However, when comes to real-world deployment, several key challenges are introduced, which complicate deployment and limit its effectiveness in real-world applications. 

In the following, we discussed the constraints from the perspective of latency, heterogeneity nature, and the computational overhead, in existing LLM collaboration and refinement scenarios.
\begin{itemize}
    \item The collaborative process across LLM inference and refinement inherently incurs communication and coordination overhead. This characteristic inevitably increases response latency, which serves as a critical metric for quality of service in task-oriented communications. Even when collaboration occurs locally across on-device models, the sequential or iterative nature of multi-model inference, query decomposition, and result aggregation or summarization can accumulate delays that negate the latency advantages of edge deployment, particularly for time-sensitive tasks where near-real-time or latency response are expected.
    \item Tailored by the heterogeneous capacity of devices in edge networks, the naturally heterogeneity of model architectures presents fundamental compatibility barriers that preclude efficient knowledge sharing mechanisms. Unlike homogeneous ensemble methods where models can directly exchange weights or gradients (i.e., federated learning), collaborating LLMs with different tokenization schemes, embedding dimensions, attention mechanisms, and architectural paradigms cannot seamlessly share learned knowledge without computationally cost-expensive fusion and aggregation, thereby limiting the flexibility and scalability of collaboration across heterogeneous LLMs.
    \item Last by not least, maintaining and updating collaborative LLM systems (i.e., model aggregation, fine-tuning) imposes substantial computational and communication costs that scale poorly as the number of participating models increased. Continuous fine-tuning or adaptation through collaboration requires either expensive distributed training protocols that aggregate updates across models, or distillation pipelines that transfer knowledge between heterogeneous architectures, both of which consume significant resources and may require frequent model redeployment. The overhead will undermine the scalability and sustainability of the collaboration frameworks.
\end{itemize}

The aforementioned challenges significantly impact the performance, efficiency, and scalability of LLM collaboration frameworks, especially in resource-constrained edge networks. Existing work fails to provide a scalable solution that can simultaneously satisfy the task-dependent QoS requirements of LLM communication and collaboration.
Considering the challenges in LLM collaborations and the fact that research in this area is still in its infancy, we seek to provide a new perspective for efficient and scalable LLM collaboration.

A few key remarks on the proposed paradigm and how it differs from existing works are summarized as follows.
\begin{itemize}
    \item \textbf{Task-oriented Communication.} \textit{Federated Inference with Refinement} could adaptively change the communication schemes (i.e., cache communication and token communication) in a LLM-native manner, achieving desirable performance under different QoS requirements.
    \item \textbf{Heterogeneity Compatibility.} \textit{Federated Inference with Refinement} does not have homogeneity assumption on model architectures, which provides a scalable solution for heterogeneous model collaboration over edge networks.
    \item \textbf{Bi-directional Collaboration.} The proposed paradigm enable bi-directional communication and collaboration, enabling a new perspective that smaller model could also has the potential to refine the inference performance of larger model via cache communications. Detailed validation results could be found in the following case study.
    \item \textbf{Lightweight Refinement Cost.} The proposed paradigm refines the LLM inference performance without requiring transmitting or updating model weights. 
\end{itemize}

\end{document}